\documentclass[aps,prl,twocolumn,showpacs,groupedaddress]{revtex4}
\usepackage[dvips]{graphicx}
\begin{document}

\title{Self-excitation of surface plasmon polaritons}

\author{V.G.~Bordo}
\email{bordo@mci.sdu.dk}

\affiliation{NanoSyd, Mads Clausen Institute, Syddansk Universitet, Alsion 2, DK-6400 S{\o}nderborg, Denmark}


\date{\today}

\begin{abstract}
The novel effect of self-excitation of surface plasmons (SESP) in a plasmonic nanocavity is predicted and its theory is developed from first principles. It is assumed that the cavity is formed by a nanogap between two metals and contains polarizable inclusions. Basing on the dyadic Green's function of the structure, the equations for the field in the cavity are investigated. It is shown that under certain conditions the field becomes unstable that leads to its self-excitation. The threshold criterion for self-excitation as well as the frequency of self-oscillation are derived in an analytical form. The SESP effect is explained in terms of a positive feedback for the polarization of inclusions provided by the field reflected from the cavity walls. Such a mechanism does not imply stimulated emission that distinguishes it from SPASER or plasmon laser.
\end{abstract}


\maketitle


Surfaces of metals and their interfaces with dielectrics support electromagnetic excitations known as surface plasmon polaritons (SPPs), or briefly surface plasmons (SPs). Their field is localized near the boundary in a narrow region of a thickness which is less than the optical radiation wavelength. This remarkable property has been used in plenty of metal nanostructures to generate, control and manipulate electromagnetic fields on a nanoscale that is the subject matter of plasmonics \cite{Mayer07}. This field holds promise for developing diverse optical and optoelectronic nanodevices for subwavelength waveguiding \cite{Bozhevolnyi10}, light energy concentration \cite{Schuller10}, ultra-sensitive sensing \cite{Zayats09}, high-resolution microscopy \cite{Novotny03}, ultra-fast computations and many other applications. \\
One of the most serious obstacles preventing from wide utilization of SPPs in photonic circuits is high metallic losses which lead to short SPP propagation lengths in the optical spectral range. Moreover, the SPP mode losses increase with the mode confinement that blocks the further miniaturization of plasmonic devices. As a solution to this problem, it has been suggested to introduce optical gain (i.e. a population inversion) in the dielectric material bordering the metal surface to reduce SPP propagation losses \cite{Sudarkin89}. This effect has been demonstrated experimentally for simple plasmonic waveguides \cite{Zhang08,Dereux09}. It was also proposed that in such systems Surface Plasmon Amplification by Stimulated Emission of Radiation (SPASER) can occur if the gain medium undergoes a strong enough pumping \cite{Stockman03,Stockman11}. This mechanism can be implemented in plasmonic metamaterials which are considered as being promising in the field of active nanoplasmonics \cite{Hess12}. \\
However, currently available gain materials put limits for possible spaser characteristics, such as spasing threshold and output intensity, and for its operation conditions \cite{Leosson12}. Moreover, for the tightly confining sub-wavelength plasmonic waveguides with electrically pumped semiconductor gain medium the current densities required to achieve SPP amplification and spasing are unsustainably high \cite{Sun12,Sun12a}. These limitations can be weakened to some extent in a three-level scheme of pumping with one of the transitions coupled to a plasmon mode \cite{Scully13}. In such a case, spasing can be achieved without population inversion on the spasing transition. This scheme nevertheless requires pumping at the other two transitions one of which should be coherently driven.\\
In this Letter, we propose a principally new avenue to SPP generation. It is based on the utilization of a feedback mechanism which exists for an oscillating dipole moment in close vicinity (within a wavelength) from a reflecting surface. Such a dipole is driven by its own reflected field that modifies essentially its dynamics \cite{Chance74}. In particular, its damping constant is renormalized in accordance with the imaginary part of the reflected field at the dipole position. In the case where the latter quantity is negative the reflected field performs a positive work on the dipole oscillations thus providing a positive feedback. This effect can occur, in principle, in any subwavelength structures, for example, between two parallel surfaces \cite{Dutra96,Nha96}, inside or nearby a cylindrical waveguide \cite{Nha97}, or inside a cylindrical cavity \cite{Bordo12}. If an ensemble of dipoles is located nearby a surface their mutual interaction mediated by the reflected field can cause a collective instability which is accompanied by the ensemble polarization amplification \cite{Bordo13}. \\
In the case of a dipole in/at a metal cavity the dipole field launches SPP modes \cite{Yip70} which act back on the dipole polarization thus establishing a feedback loop.  The considered mechanism is an essentially classical effect. It does not imply stimulated emission in the structure and hence it does not require a population inversion in the dipole. In this sense, the predicted here effect of Self-Excitation of Surface Plasmons (SESP) is principally different from SPASER or plasmon laser \cite{Ma13}. \\
The proposed principle has much in common with the other processes which are known in literature as self-excited oscillations or self-oscillations \cite{Jenkins13}. All such phenomena are closed loop processes which employ a positive feedback. Their point of departure is a small initial deviation or field (a seed) leading to an unstable initial exponential increase of the output. This increase is usually limited by a certain mechanism, for example by saturation. Lasers and spasers are another examples of such systems where a seed is provided by the initial number of spontaneously emitted quanta.\\
In this Letter, we consider the SESP effect in a metal-insulator-metal (MIM) cavity. Basing on the dyadic Green's function of this structure we derive a criterion for self-excitation of SPPs as well as an expression for the frequency of self-oscillation. \\
\begin{figure}
\includegraphics[width=80mm]{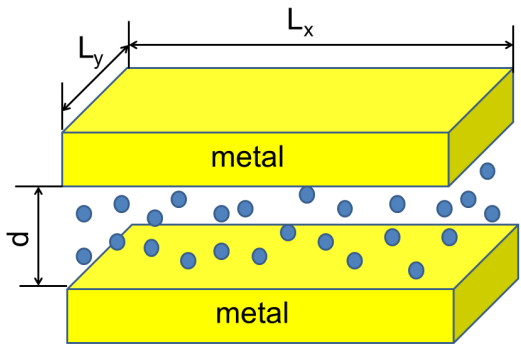}
\caption{\label{fig:geometry}(color online). Geometry of the MIM cavity under consideration. Inclusions in the gap material are shown by dark blue circles. The cavity is enclosed by perfectly conducting walls (not shown).}
\end{figure}
Let us consider a sub-wavelength gap between two identical metals with the dielectric function $\epsilon_m(\omega)$ which is enclosed from all sides with perfectly conducting plates \cite{Note1} (see Fig. \ref{fig:geometry}). Let us assume that the gap (host) material has the dielectric function $\epsilon_h$ and contains polarizable inclusions which have a resonance at the frequency $\omega_0$. Let the gap thickness be $d$ and its two other extensions be $L_x$ and $L_y$ along the $x$ and $y$ axes, respectively. Let us assume that such a cavity undergoes excitation by an external electromagnetic field which is given by ${\bf E}_i({\bf r})\exp(-i\omega t)$ inside the gap with $\omega$ being its frequency. We shall then seek the field in the cavity in the form $\tilde{{\bf E}}({\bf r},t)\exp(-i\omega t)$, where a tilde denotes an amplitude which varies in time much slower than $\exp(\pm i\omega t)$. \\
The external field polarizes the inclusions which in their turn launch the SPP modes of the cavity. On the other hand, the SPP field acts back on the polarization of inclusions thus providing a feedback mechanism. This effect can be described in terms of the field susceptibility tensor \cite{Sipe84}, $\bar{\bf F}$, of the system as follows 
\begin{equation}\label{eq:field}
\tilde{{\bf E}}({\bf r},t)\approx{\bf E}_i({\bf r})+\int \bar{\bf F}({\bf r},{\bf r}^{\prime};\omega)\tilde{{\bf P}}({\bf r}^{\prime},t)d{\bf r}^{\prime},
\end{equation}
where we have used the substitution ${\bf P}(t)=\tilde{{\bf P}}(t)\exp(-i\omega t)$ for the polarization of the inclusions, the integral is taken over the region occupied by the inclusions and we have neglected the effect of retardation for the field propagation from point ${\bf r}^{\prime}$ to point ${\bf r}$ \cite{Note2}. The quantity $\bar{\bf F}({\bf r},{\bf r}^{\prime};\omega)$ relates the electric field at the point ${\bf r}$ generated by a classical dipole, oscillating at frequency $\omega$, with the dipole moment itself, located at ${\bf r}^{\prime}$. It can be decomposed into direct, $\bar{\bf F}^0$, and reflected, $\bar{\bf F}^R$, contributions, the latter one being originating from the dipole field reflected from the cavity walls. \\
The polarization of inclusions can be described in the framework of the harmonic oscillator model as follows
\begin{equation}\label{eq:oscillator}
\frac{d^2{\bf P}}{dt^2}+ 2\gamma \frac{d{\bf P}}{dt}+\omega_0^2{\bf P} = a\tilde{{\bf E}}(t)e^{-i\omega t},
\end{equation}
where $\gamma$ is the relaxation constant and the coefficient $a$ characterizes the coupling between the inclusions and the field. To find the evolution of the cavity field, one has to solve Eq. (\ref{eq:oscillator}) jointly with the integral equation (\ref{eq:field}).\\
The fields can be expanded in the Fourier series over the intervals $0\le x \le L_x$ and $0\le y \le L_y$. For the amplitude $\tilde{{\bf E}}$ it has the form
\begin{equation}
\tilde{{\bf E}}({\bf r},t)=\sum_{m,n=1}^{\infty}{\bf e}_{mn}(z,t)\sin\left(\frac{\pi m}{L_x}x\right)\sin\left(\frac{\pi n}{L_y}y\right).
\end{equation}
Similar expansions take place for ${\bf E}_i$ and $\tilde{{\bf P}}$ with the coefficients ${\bf e}^i_{mn}(z)$ and ${\bf p}_{mn}(z,t)$, respectively. Here the set of integers $m$ and $n$ distinguishes between different Fabry-P{\'e}rot modes of the plasmonic cavity. In what follows, we shall restrict ourselves by the consideration of the Fabry-P{\'e}rot modes which originate from the gap SPP of the MIM waveguide \cite{Bozhevolnyi07}. Such a SPP has a transverse magnetic (TM) polarization and exists for all values of the gap thickness, $d$.\\
The substitution of the Fourier series into Eqs. (\ref{eq:field}) and (\ref{eq:oscillator}) on the assumption that $L_x,L_y\gg 2d$ leads to the following set of equations for the mode $\{m,n\}$ field:
\begin{eqnarray}\label{eq:FP1}
{\bf e}_{mn}(z,t)\nonumber \\
\approx{\bf e}_{mn}^i(z)+\int_{-d/2}^{d/2}\bar{\mathcal{F}}(z,z^{\prime};\omega,\kappa){\bf p}_{mn}(z^{\prime},t)dz^{\prime},
\end{eqnarray}
\begin{eqnarray}\label{eq:FP2}
\frac{d^2{\bf p}_{mn}(z,t)}{dt^2}+2(\gamma - i\omega)\frac{d{\bf p}_{mn}(z,t)}{dt}\nonumber\\
+(\omega_0^2-\omega^2-2i\gamma\omega){\bf p}_{mn}(z,t)=a{\bf e}_{mn}(z,t),
\end{eqnarray}
where
\begin{equation}\label{eq:kappa0}
\kappa=\pi\sqrt{\left(\frac{m}{L_x}\right)^2+\left(\frac{n}{L_y}\right)^2}
\end{equation}
and $\bar{\mathcal{F}}$ is the Fourier transform of the field susceptibility tensor. \\
To investigate the stability of the mode field one has to consider Eqs. (\ref{eq:FP1}) and (\ref{eq:FP2}) for small deviations $\delta{\bf e}_{mn}$ and $\delta{\bf p}_{mn}$. Assuming that there is no initial polarization of inclusions and performing the Laplace transform of those equations in time one comes to the integral equation 
\begin{eqnarray}\label{eq:integral}
\delta\tilde{{\bf e}}_{mn}(z,s)\nonumber\\
=\chi^{\prime}(s)\int_{-d/2}^{d/2}\bar{\mathcal{F}}^R(z,z^{\prime};\omega,\kappa)\delta\tilde{{\bf e}}_{mn}(z^{\prime},s)dz^{\prime},
\end{eqnarray}
where $\chi^{\prime}(s)=\chi(s)/[1-(4\pi/3\epsilon_h)\chi(s)]$, $\chi(s)=a/[(s-i\omega)^2+2\gamma(s-i\omega)+\omega_0^2]$ is the Laplace transform of the linear susceptibility of the inclusions which is assumed to be isotropic, $s=\sigma+i\Omega$ and the tilde above ${\bf e}_{mn}$ denotes its Laplace transform. Let us note that the quantity $\chi^{\prime}(s)$ represents the Laplace transformed linear susceptibility renormalized because of the Lorentz local field of the inclusions \cite{Born70}. The kernel $\bar{\mathcal{F}}^R(z,z^{\prime})$ in this equation is degenerate that allows one to reduce Eq. (\ref{eq:integral}) to a set of four algebraic equations \cite{Polyanin}. The condition that the determinant of this set of equations, $D(s)$, is equal to zero determines the values of $s$ for which nontrivial solutions exist. If there is at least one root for which $\sigma>0$ the system is unstable and any initial deviation in the mode field will lead to its exponential growth with time, i.e. self-excitation.\\
As one can show, the equation $D(s)=0$ has a root $s_0=-\gamma+i\Delta+i\beta\mathcal{F}_0^{\prime}$,
where $\mathcal{F}_0^{\prime}=\mathcal{F}_0+(4\pi/3\epsilon_h)$ and
\begin{eqnarray}
\mathcal{F}_0=\frac{4\pi idR_pe^{iW_gd}}{1-R_p^2e^{2iW_gd}}\left[W_g\left(R_pe^{iW_gd}-\frac{\sin W_gd}{W_gd}\right)\right.\nonumber\\
\left.+\frac{\kappa^2}{W_g}\left(R_pe^{iW_gd}+\frac{\sin W_gd}{W_gd}\right)\right],
\end{eqnarray}
with $W_g=\sqrt{(\omega/c)^2\epsilon_h-\kappa^2}$, $c$ the speed of light in vacuum, $R_p$ the Fresnel reflection coefficient for $p$-polarized light at the gap-metal interface taken for the wavevector component along the interface equal to $\kappa$, $\Delta=\omega-\omega_0$ and $\beta=a/(2\omega_0)$.\\
The real part of $s_0$ determines the criterion for the mode self-excitation:
\begin{equation}\label{eq:generation}
\text{Re}(\chi_r\mathcal{F}_0)>1,
\end{equation}
where we have introduced the linear susceptibility of inclusions at the resonance $\Delta=0$, $\chi_r=i\beta/\gamma$. On the other hand, $\text{Im}(s_0)\equiv \Omega_0$, gives the frequency of the mode self-oscillation, $\omega_{so}$:
\begin{equation}\label{eq:frequency}
\omega_{so}=\omega-\Omega_0=\omega_0-\gamma\text{Im}(\chi_r\mathcal{F}_0^{\prime}),
\end{equation}
which differs from both $\omega$ and $\omega_0$. Let us note that the criterion (\ref{eq:generation}) can be obtained from Poynting's theorem. It can be interpreted as a condition that there is a net inward flux of the electromagnetic energy, scattered by the inclusions and reflected back into the gap, which is larger than the rate of the energy dissipation in the inclusions.\\
An additional rule for a non-zero mode field follows from Eq. (\ref{eq:FP1}). The exciting field should have a non-zero Fourier component ${\bf e}^i_{mn}$, which in turn should have a non-zero projection onto the plane containing both the $z$ axis and the mode wave vector, $\vec{\kappa}_{mn}=(\pi m/L_x,\pi n/L_y)$ (we imply a TM polarization of the mode).\\
In the self-oscillation criterion (\ref{eq:generation}), the quantity $\mathcal{F}_0$ is determined completely by the properties of the cavity, whereas the parameter $\chi_r$ depends solely on the properties of the inclusions. Up to this point, we have not specified anywhere the physical origin of the polarizable inclusions. They can be represented by impurity atoms, ions, molecules, quantum dots (QDs) or metal nanoparticles (NPs). For a two-level model of inclusions, that can be applied for atomic systems or QDs, $\chi_r$ is found as $i\mu^2N /(\hbar \gamma_{\perp} )$ with $\mu$ the root mean square of the dipole transition matrix element, $\gamma_{\perp }$ the transverse relaxation rate and $N$ the number density of inclusions which are supposed to be in the ground state. If the gap material is represented by a nanocomposite which contains spherical metal NPs, one can estimate $\chi_r$ as $3i\epsilon_hf\omega_{sp}/(4\pi\Gamma)$ with $f$ the volume fraction of NPs, $\omega_{sp}$ the localized surface plasmon polariton (LSPP) frequency and $\Gamma$ the electron collision rate in the NP material.  \\
The approach developed above predicts an exponential growth of the mode field with time. This is valid, however, only upon condition that the saturation effect is negligible, i.e. as long as the nonlinear terms in Eq. (\ref{eq:oscillator}), which describe the evolution of the  polarization, can be omitted. For metal nanoparticles in a nanocomposite this effect becomes essential at very high field intensities of the order of 50 GW/cm$^2$ \cite{Plaksin08}.\\
Let us note that the theory developed here is formulated in completely classical terms. On the other hand, it has much in common with the semiclassical laser (or spaser) theory where stimulated emission plays key role. This similarity is, however, not accidental. According to papers \cite{Cray82,Fain87} stimulated emission can be described classically if one introduces an appropriate phase shift between the dipole moment and the driving force in Eq. (\ref{eq:oscillator}). Usually this effect disappears for an unperturbed (non-excited) system due to the ensemble average over phases. However in our case all inclusion dipoles have the same phase with respect to their own reflected field in virtue of the condition $d\ll \lambda =2\pi c/\omega$.\\
\begin{figure}
\includegraphics[width=80mm]{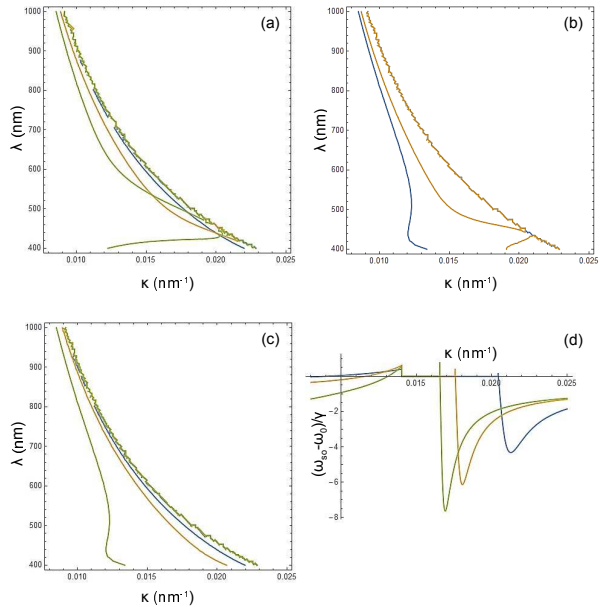}
\caption{\label{fig:islands}(color online). Results of calculations for the MIM cavity composed of a glass slab ($\epsilon_h=1.45^2$) containing Ag NPs ($f=0.02$) between two metal surfaces. (a)-(c) Islands of instability. (a) A slab of different thickness is enclosed between Ag surfaces: $d=50$ nm (blue curve), $d=100$ nm (brown curve) and $d=150$ nm (green curve). $\Gamma=4.99\cdot 10^{14}$ s$^{-1}$. (b) A slab of thickness $d=50$ nm is enclosed between different metals: Ag (blue curve) and Au (brown curve). $\Gamma=1.72\cdot 10^{14}$ s$^{-1}$. (c) A slab of thickness $d=50$ nm containing NPs with different $\Gamma$ is enclosed between two Ag surfaces (the corresponding NP radius is given in brackets): $\Gamma=4.99\cdot 10^{14}$ s$^{-1}$ ($R=3$ nm) (blue curve), $\Gamma=3.12\cdot 10^{14}$ s$^{-1}$ ($R=5$ nm) (brown curve) and $\Gamma=1.72\cdot 10^{14}$ s$^{-1}$ ($R=10$ nm) (green curve). (d) Frequency pulling of the mode self-oscillation for $\lambda =650$ nm as a function of $\kappa$.  A slab of different thickness is enclosed between Ag surfaces: $d=50$ nm (blue curve), $d=100$ nm (brown curve) and $d=150$ nm (green curve). $\Gamma=4.99\cdot 10^{14}$ s$^{-1}$. The discontinuities in the plots correspond to the regions where the series for the field susceptibility given in Ref. \cite{Nha96} do not converge.}
\end{figure}
We illustrate the general theory developed above by some numerical calculations. We assume that the gap is filled with glass which contains spherical Ag NPs. We adopt the Drude model for the dielectric function of metals, both in the cladding and in NPs, $\epsilon_m(\omega)=\epsilon_{\infty}-\omega_p^2/[\omega(\omega+i\Gamma)]$, where $\omega_p$ is the plasma frequency, $\Gamma$ is the relaxation rate and $\epsilon_{\infty}$ is the offset which takes into account the interband transitions \cite{Shalaev10}. For NPs the quantity $\Gamma$ includes the size effect contribution and can be written as $\Gamma=\Gamma_{\infty}+av_F/R$, where $\Gamma_{\infty}$ is the damping constant for an unbounded metal, $a\approx 1$, $v_F$ is the Fermi velocity and $R$ characterizes the size of the metal particle \cite{Shalaev10}. Using the parameters given in Ref. \cite{Shalaev10} for silver, one obtains $\lambda_{sp}\equiv 2\pi c/\omega_{sp}\approx 410$ nm.\\
Figures \ref{fig:islands}(a)-(c) show the contour plots $\text{Re}(\chi_r\mathcal{F}_0)=1$ in the plane $\kappa$ - $\lambda$. The chosen values of the NP sizes and volume fraction correspond to those which can be obtained for Ag NPs embedded in a glass matrix \cite{Agarwal15,Stalmashonak}. The areas embraced by these contours correspond to the values of $\kappa$ and $\lambda$ for which the mode field is unstable. They can thus be called the 'islands of instability'. Here $\lambda$ is the wavelength of the exciting field whereas $\kappa$ is defined by Eq. (\ref{eq:kappa0}) and can be varied by changing the lengths $L_x$ and $L_y$. The kinks at $\lambda\approx 410$ nm originate from the resonance with LSPPs in metal NPs. All islands have a common upper-right border which is described by the equation $\lambda=2\pi\sqrt{\epsilon_h}/\kappa$ or, equivalently, $W_g=0$. The vicinity of this curve is favorable for a feedback provided by the surface reflections: It corresponds to a grazing incidence of the waves, scattered by NPs, onto the cladding surfaces for which $R_p\approx -1$ and hence the losses in the metal cladding are minimized. As one can see, the area of the islands increases with the increase of $d$. It increases also with the decreasing of $\Gamma$ (i.e. with the increasing of the NP size). Figure \ref{fig:islands}b illustrates how the shape of the islands depends on the metal cladding. The behavior of the frequency pulling of self-oscillation, $(\omega_{so}-\omega_0)/\gamma$, is shown in Fig. \ref{fig:islands}d. The parts of this dependence which are outside the instability regions depicted in Fig. \ref{fig:islands}a correspond to decaying self-oscillations. This effect depends noticeably on the gap thickness, but depends only slightly on the cladding metal (not shown).\\
In conclusion, we have developed here the theory of SPP self-excitation in the MIM structure from first principles. We have derived an analytical criterion for the SPP self-excitation, Eq. (\ref{eq:generation}), and an expression for the frequency pulling effect, Eq. (\ref{eq:frequency}). For a given cavity, the self-excitation threshold condition is imposed on the inclusions linear susceptibility, $\chi_r$, rather than on the population inversion as in conventional lasers or spasers. The spectral interval where the SESP effect can occur spans over the visible and infrared regions and can be engineered by a proper design of the MIM cavity. An additional opportunity for tuning the resonances of the linear susceptibility $\chi$ can be provided by a choice of the metal composition, size and shape of NPs as inclusions in the gap material \cite{Lee06}. For example, for a rod-shaped NP the LSPP resonance splits into two, one of which can be shifted to regions where the optical losses are lower \cite{Pelton08}. The general character of the polarization loop gain discussed here suggests that the SESP effect is not specific for the considered structure and can be observed in other subwavelength hybrid plasmonic structures. Its experimental verification will open up a prospect for the development of a new generation of nanoscale active plasmonic devices which do not require a powerful pumping and hence will not be highly energy-consuming.

\end{document}